\documentclass[prd,twocolumn,a4paper,superscriptaddress,floatfix]{revtex4}
\usepackage{graphicx}
\usepackage{bbm}
\usepackage[all]{xy}
\usepackage{amsmath}
\usepackage{amssymb}
\usepackage{epstopdf}

\def\be{\begin{equation}}
\def\ee{\end{equation}}
\newcommand{\bq}{\begin{eqnarray}}
\newcommand{\eq}{\end{eqnarray}}
\newcommand{\bes}{\begin{subequations}}
\newcommand{\ees}{\end{subequations}}

\def\ben{\begin{eqnarray}}
\def\een{\end{eqnarray}}
\def\ba{\begin{array}}
\def\ea{\end{array}}

\begin{document}
\newcommand{\half}{{\textstyle\frac{1}{2}}}
\allowdisplaybreaks[3]
\def\a{\alpha}
\def\b{\beta}
\def\g{\gamma}\def\G{\Gamma}
\def\d{\delta}\def\D{\Delta}
\def\ep{\epsilon}
\def\et{\eta}
\def\z{\zeta}
\def\t{\theta}\def\T{\Theta}
\def\l{\lambda}\def\L{\Lambda}
\def\m{\mu}
\def\f{\phi}\def\F{\Phi}
\def\n{\nu}
\def\p{\psi}\def\P{\Psi}
\def\r{\rho}
\def\s{\sigma}\def\S{\Sigma}
\def\ta{\tau}
\def\x{\chi}
\def\o{\omega}\def\O{\Omega}
\def\k{\kappa}
\def\pa {\partial}
\def\ov{\over}
\def\br{\\}
\def\ud{\underline}

\newcommand\lsim{\mathrel{\rlap{\lower4pt\hbox{\hskip1pt$\sim$}}
    \raise1pt\hbox{$<$}}}
\newcommand\gsim{\mathrel{\rlap{\lower4pt\hbox{\hskip1pt$\sim$}}
    \raise1pt\hbox{$>$}}}
\newcommand\esim{\mathrel{\rlap{\raise2pt\hbox{\hskip0pt$\sim$}}
    \lower1pt\hbox{$-$}}}
\newcommand{\dpar}[2]{\frac{\partial #1}{\partial #2}}
\newcommand{\sdp}[2]{\frac{\partial ^2 #1}{\partial #2 ^2}}
\newcommand{\dtot}[2]{\frac{d #1}{d #2}}
\newcommand{\sdt}[2]{\frac{d ^2 #1}{d #2 ^2}}    
\newcommand{\aap}{A {\&}A}
\newcommand{\mnras}{MNRAS}

\title{Generalized Layzer-Irvine equation: the role of dark energy perturbations in cosmic structure formation}

\author{P. P. Avelino}
\email[Electronic address: ]{pedro.avelino@astro.up.pt}
\affiliation{Centro de Astrof\'{\i}sica da Universidade do Porto, Rua das Estrelas, 4150-762 Porto, Portugal}
\affiliation{Departamento de F\'{\i}sica e Astronomia da Faculdade de Ci\^encias
da Universidade do Porto, Rua do Campo Alegre 687, 4169-007 Porto, Portugal}
\author{C. F. V. Gomes}
\email[Electronic address: ]{claudio.gomes@astro.up.pt}
\affiliation{Centro de Astrof\'{\i}sica da Universidade do Porto, Rua das Estrelas, 4150-762 Porto, Portugal}
\affiliation{Departamento de F\'{\i}sica e Astronomia da Faculdade de Ci\^encias
da Universidade do Porto, Rua do Campo Alegre 687, 4169-007 Porto, Portugal}

\begin{abstract}

We derive, using the spherical collapse model, a generalized Layzer-Irvine equation which can be used to describe the gravitational collapse of cold dark matter in a dark energy background. We show that the usual Layzer-Irvine equation is valid if the dark matter and the dark energy are minimally coupled to each other and the dark energy distribution is homogeneous, independently of its equation of state. We compute the corrections to the standard Layzer-Irvine equation which arise in the presence of dark energy inhomogeneities. We show that, in the case of a dark energy component with a constant equation of state parameter consistent with the latest observational constraints, these corrections are expected to be small, even if the dark energy has a negligible sound speed. However, we find that, in more general models, the impact of dark energy perturbations on the dynamics of clusters of galaxies, which will be constrained by ESA's Euclid mission with unprecedented precision, might be significant.

\end{abstract} 
\maketitle

\section{Introduction}

The Layzer-Irvine (LI) equation \cite{Irvine:1961,Layzer:1963}, also known as cosmic energy equation, describes the dynamics of local dark matter perturbations in an otherwise homogeneous and isotropic universe. It has been used in determinations of the matter density, cluster mass and size, and the galaxy peculiar velocity field \cite{Davis:1997vg,Liddle,Fukugita:2004ee,Zaroubi:2004vu} and, more recently, as a crucial test to the accuracy of cosmological N-body simulations in the non-linear regime \cite{Joyce:2013,Labini:2013}. 

In its original form, the LI equation accounts for the evolution of the energy of a system of non-relativistic particles, interacting only through gravity, until virial equilibrium is reached, but it has recently been generalized to account for a non-minimal interaction between dark matter and a homogeneous dark energy (DE) component  \cite{Bertolami:2007zm,Abdalla:2007rd,Abdalla:2009mt,Bertolami:2012yp,Avelino:2011zx} (see also \cite{Shtanov:2010iy} for a generalization of the LI equation to modified gravity scenarios). A deviation from the usual virial relation in galaxy clusters is expected as a result of such an interaction \cite{Bertolami:2007zm,Abdalla:2007rd,Abdalla:2009mt,Bertolami:2012yp,Avelino:2011zx}  and its observational detection would be a key step in the search for the nature of dark matter and DE.

In quintessence models, DE is characterized by a sound speed which is equal to the speed of light in vacuum. Hence, the DE fluctuations associated to the gravitational collapse of matter perturbations are necessarily very small on cosmological scales \cite{Mota:2007zn,Avelino:2008cu}. However, this does not have to be the case in more general models \cite{Bean:2003fb,Nunes:2004wn,Nunes:2005fn,Abramo:2007iu,Ballesteros:2008qk,Abramo:2009ne,Creminelli:2009mu,dePutter:2010vy,Blomqvist:2010ky} and, consequently, it is reasonable to expect that DE perturbations could play a relevant role in the dynamics of galaxy clusters. 

In this paper, our main goal is to generalize the LI equation to account for the presence of DE perturbations. We start in Sec. \ref{layz} by  presenting the standard LI equation. Then, in Sec. \ref{sphe}, we use the spherical collapse model to determine the evolution of the (peculiar) gravitational and kinetic energies associated to the cold dark matter (CDM) inhomogeneities. In Sec. \ref{gene} we generalize the LI equation to account for DE perturbations, quantifying the departures from the standard case in various scenarios consistent with current data. Finally, we conclude in Sec. \ref{conc}.

\section{Standard Layzer-Irvine Equation\label{layz}}

Consider a local inhomogeneity associated to $N$ point mass CDM particles of mass $m_{[j]}$, whose trajectories are given by ${\bf r}_{[j]}=a(t) {\bf x}_{[j]}$ with $j=1,\ldots,N$ ($a$ is the scale factor and $ {\bf x}_{[j]}$ represents the comoving position of the particles). The Hamiltonian for this system can be written as \cite{Peebles}
\be
{\mathcal E}={\mathcal K}+{\mathcal U}\,,
\ee
where
\bq
{\mathcal K}&=& \sum_{j=1} ^N\frac{p_{[j]}^2}{2m_{[j]}}\,,\label{K}\\
{\mathcal U}&=&-\frac{G}{2}\int \frac{\left[\rho_m({\bf r})-{\bar \rho}_m\right]  \left[\rho_m({\bf r}')-{\bar \rho}_m\right]}{|{\bf r}-{\bf r}'|} d^3 {\bf r} d^3 {\bf r}'\label{U}\,,
\eq
are, respectively, the total (peculiar) kinetic and gravitational potential energy, $G$ is the gravitational constant, $p_{[j]}=|{\bf p}_{[j]}|$, ${\bf p}_{[j]}=m_{[j]} {\bf v}_{[j]}$, ${\bf v}_{[j]}={\dot {\bf r}}_{[j]}-H {\bf r}_{[j]}=a {\dot {\bf x}}_{[j]}$ is the peculiar velocity of the CDM particles, $v_{[j]}=|{\bf v}_{[j]}|$, a dot represents a total derivative with respect to the physical time $t$, and ${\bar \rho}_m$ is the average value of the matter density $\rho_m$.  The classical energy equation is
\be
{\dot {\mathcal E}} \equiv \frac{d\mathcal{E}}{dt}= \frac{\partial \mathcal{E}}{\partial t} \label{eq:energy equation},
\ee
where the partial derivative with respect to the physical time is computed at fixed particle comoving coordinates ${\bf x}_{[j]}$ and comoving momenta ${\bf p}_{[j]}/a=m_{[j]} {\dot {\bf x}}_{[j]}$. This way, one has  ${\mathcal U} \propto a^{-1}$ and ${\mathcal K} \propto a^{-2}$. Consequently, using Eq. (\ref{eq:energy equation}) one finally obtains
\be
{\dot {\mathcal E}}+ H(2 {\mathcal K}+{\mathcal U})=0 \,, \label{eq:LI1}
\ee
where $H={\dot a}/a$. This is the standard LI equation which is valid throughout the entire process of structure formation both in the linear and non-linear regimes. The virial equation, ${\mathcal K}=-{\mathcal U}/2$, holds in the case of relaxed non-linear objects with ${\dot {\mathcal E}} = 0$.

\section{Spherical Collapse Model \label{sphe}}

Consider two homogeneous concentric spherical patches, whose dynamics are described by the scale factors $a_1$ (backround patch) and $a_2$ (perturbed patch) and assume that the total mass in CDM particles is conserved or, equivalently, that CDM and DE are minimally coupled. The peculiar velocity of the CDM particles at the (perturbed) position ${\bf r}_{2[j]}$, with respect to the center of the patches, is given by
\be
{\bf v}_{pec}({\bf r}_{2[j]})=\Delta H {\bf r}_{2[j]}=a_2 \Delta H{\bf q}_{[j]}\,,
\ee
where ${\bf q}_{[j]}={\bf r}_{2[j]}/a_2$ represents the comoving position of the CDM particles, $\Delta H \equiv H_2-H_1$ and the subscripts $1$ and $2$ refer to the background and perturbed patches, respectively. The total (peculiar) kinetic energy associated with the spherical inhomogeneity of comoving size $q=|{\bf q}|$ can be computed as
\bq
\frac{\mathcal K}{M}&\equiv & \frac12 \langle v_{pec}^2 \rangle= \frac12 \frac{\int_0^q v_{pec}^2(q') {q'}^2 dq' }{\int_0^q {q'}^2 dq'}=\\
&=&\frac{3}{10} (a_2 \Delta H)^2   q^2\,,
\eq
where 
\be
M=\frac{4\pi}{3}\rho_{m2} r_2^3=\frac{4\pi}{3}\rho_{m1} r_1^3\,,
\ee
$r_1=a_1q$ and $r_2=a_2 q$. The total mass $M$ is conserved and, consequently, $\rho_{m1} \propto a_1^{-3}$ and $\rho_{m2} \propto a_2^{-3}$. The density perturbation of the CDM component and its time derivative are 
\bq
\delta &\equiv& \frac{\rho_{m2}-\rho_{m1}}{\rho_{m1}}=\left(\frac{a_1}{a_2}\right)^3-1\,,\label{delta}\\
{\dot \delta} &=& -3 \left(\frac{a_1}{a_2}\right)^3 \Delta H\label{deltad}\,.
\eq
Consequently, specifying initial conditions for $a_1$, $a_2$, $H_1$ and $H_2$ is enough to define the initial values of $\delta$ and $\dot \delta$ (note that $\Delta H = -{\dot \delta}/(3(\delta+1))$.

The unperturbed matter density is given by 
\be
\rho_{m1}=\frac{3H_1^2 \Omega_{m1}}{8\pi G}\,,
\ee
where $\Omega_{m}=\rho_{m}/\rho_{c}$ is the fractional matter density parameter and $\rho_c \equiv 3 H^2/(8\pi G)$ is the critical density. In this paper we shall use time units with $8 \pi G \rho_{m1i}/3=1$ (or, equivalently, $H_{1i}^2 \Omega_{m1i}=1$, where the subscript `$i$' represents some early initial time deep in the matter dominated era). Making also the choice of scale factor normalization $a_{1i}=1$ one obtains 
\be
M=\frac{4\pi}{3}\rho_{m1i} q^3=\frac{4\pi}{3}\rho_{m2i} \left(a_{2i} q\right)^3=\frac{q^3}{2G}\,,
\ee
with $8 \pi G \rho_{m1}/3=a_1^{-3}$.
On the other hand, the perturbed mass density can be written as
\be
\rho_{m2}=\frac{3H_2^2 \Omega_{m2}}{8\pi G}=\rho_{m1}\left(\frac{a_1}{a_2}\right)^3\,,
\ee
so that $8 \pi G \rho_{m2i}/3=H_{2i}^2 \Omega_{m2i}=a_{2i}^{-3}$ and  $8 \pi G \rho_{m2}/3=a_2^{-3}$.

The (peculiar) gravitational energy of the CDM particles may be computed using Eq. (\ref{U}). The result is given by
\be
{\mathcal U}={\mathcal U}_A+{\mathcal U}_B+{\mathcal U}_C\,,
\ee
where
\bq
{\mathcal U}_A&=&-\frac{3}{5}\frac{GM_{+}^2}{r_2}\,, \\
{\mathcal U}_B&=&-\frac{3}{2}\frac{GM_{-}^2}{r_1}\left(1-\left(\frac{r_2}{r_1}\right)^2\right) \times \nonumber\\
&\times& \left(\frac{M_{+}}{M_{-}}-\left(\frac{r_2}{r_1}\right)^3\right)\,, \\
{\mathcal U}_C&=&-\frac{3}{5}\frac{GM_{-}^2}{r_1} \left(1-\left(\frac{r_2}{r_1}\right)^5\right)\,,
\eq
with
\be
M_{+}=\frac{4\pi}{3}\rho_{+} r_2^3\,, \qquad M_{-}=\frac{4\pi}{3}\rho_{-} r_1^3\,,
\ee
$\rho_{+}=\rho_{m2}-\rho_{m1}$ and $\rho_{-}=-\rho_{m1}$.

Defining $U=U_A+U_B+U_C$ with
\be
E=\frac{G}{q^5}\,{\mathcal E}\,, \quad  K=\frac{G}{q^5\,}{\mathcal K}\,, \quad U_{A,B,C}=\frac{G}{q^5}\,{\mathcal U}_{A,B,C}\,,
\ee
one obtains
\bq
U_A&=&-\frac{3}{20}a_2^{-1}\left(1-\left(\frac{a_2}{a_1}\right)^3\right)^2\label{UA}\,, \\
U_B&=&\frac{3}{8} a_1^{-1}\left(1-\left(\frac{a_2}{a_1}\right)^2\right)\,,\\
U_C&=&-\frac{3}{20} a_1^{-1}\left(1-\left(\frac{a_2}{a_1}\right)^5\right)\,,\\
K&=&\frac{3}{20} (a_2 \Delta H)^2\label{K1}\,.
\eq
Taking the derivative with respect to time one finds
\bq
{\dot U}_A&=&\frac{3}{20} \frac{H_2}{a_2}\left(1-\left(\frac{a_2}{a_1}\right)^3\right) \times \nonumber\\
&\times& \left(1+ \left(\frac{a_2}{a_1}\right)^3\left(5 - 6 \frac{H_1}{H_2}\right)\right)\label{LI1}\,.\\
{\dot U}_B&=&-\frac{3}{8}\frac{H_1} {a_1}\left(1-\left(\frac{a_2}{a_1}\right)^2\left(3-2 \frac{H_2}{H_1}\right)\right)\,,\\
{\dot U}_C&=& \frac{3}{20}\frac{H_1} {a_1}\left(1-\left(\frac{a_2}{a_1}\right)^5\left(6-5 \frac{H_2}{H_1}\right)\right)\,,\\
{\dot K}&=&\frac{3}{10}\Delta H \left(H_2^2 -H_1 H_2+{\dot H}_2-{\dot H}_1\right) a_2^2\label{LI4}\,,
\eq

\section{Generalized Layzer-Irvine Equation \label{gene}}

The results obtained in the previous section using the spherical collapse model may be combined in a generalized LI equation which takes into account the role of inhomogeneities in the DE component. Summing Eqs.  (\ref{LI1}-\ref{LI4}) and using Eqs.  (\ref{UA}-\ref{K1}) one finally obtains
\be
{\dot E}+ H_1(2 K+U)=\frac{3}{10}\Delta H\Delta f a_2^2\label{LIg1}\,, 
\ee
where $\Delta f=f_2-f_1$ and
\be
f={\dot H}+H^2+\frac12 a^{-3} \,.
\ee
By using Eqs. (\ref{delta}-\ref{deltad}), Eq. (\ref{LIg1}) may also be written as
\be
{\dot E}+ H_1 \left((1+\alpha)2K+U\right)=0\,, 
\ee
with
\be
\alpha=-\frac{1}{H_1}\frac{\Delta f}{\Delta H}\,.
\ee

\subsection{Homogeneous Dark Energy \label{homo}}

Let us start by assuming that the DE component is roughly homogeneous so that only the CDM component is perturbed. The derivative of the Hubble parameter with respect to cosmic time can be written as
\be
{\dot H}=\frac{\ddot a}{a}-H^2\,,
\ee
where the acceleration is given by the Raychaudhury equation 
\be
\frac{\ddot a}{a}=-\frac{4\pi G}{3}\left[(1+3w)\rho_w+\rho_m\right]\,,
\ee
$\rho_w$ is the DE density and $p_w=w \rho_w$ is the DE pressure ($w$ is the DE equation of state parameter). Remembering that our choice of time units and scale factor normalization implies that $8 \pi G \rho_m/3=a^{-3}$ in both background and perturbed patches ($1$ and $2$, respectively), one obtains
\bq
f&=&{\dot H}+H^2+\frac12 a^{-3} =\frac{\ddot a}{a}+\frac{4\pi G \rho_m}{3}= \nonumber\\
&=&-\frac{4\pi G}{3}\left[(1+3w)\rho_w\right]\,.
\label{rach}
\eq
If the DE is homogeneous then $\rho_{w1}=\rho_{w2}$ and, consequently, $\Delta f=0$. This implies that usual form of the LI  equation is valid in this case, regardless of the particular form of the DE equation of state, thus confirming the result obtained in \cite{Avelino:2011zx}.

\begin{figure}
	\centering
	\includegraphics[width=8.0cm]{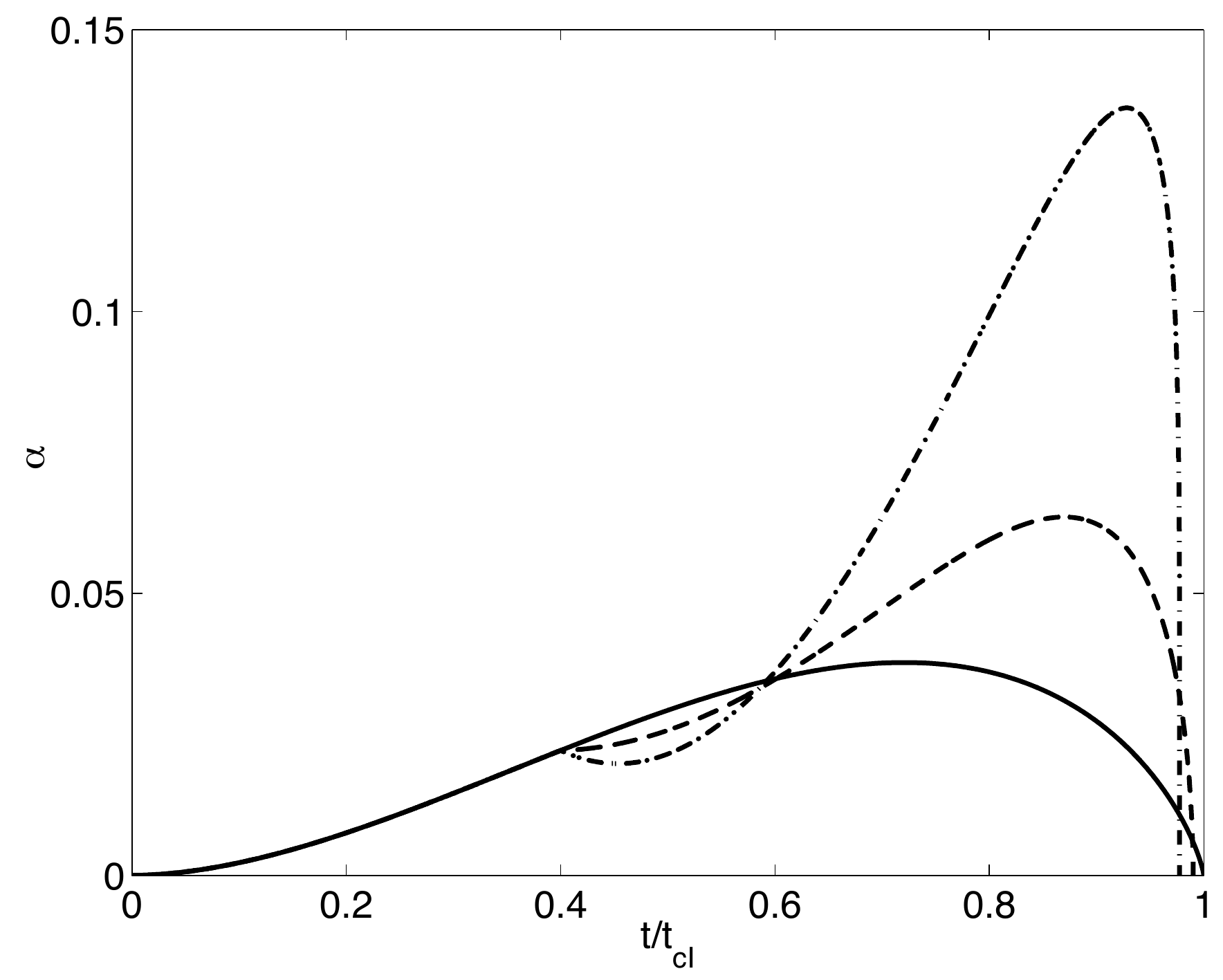}
\caption{Evolution of the parameter $\alpha$ with $t/t_{cI}$ for three different models ($t_{cI}$ is the perturbation collapsing time in model I). In all models the background value of the DE equation of state parameter is fixed at $w_1=-0.95$. In Model I (solid line) $w_2=-0.95$ in the perturbed region at all times, while in Models II (dashed line) and III (dot-dashed line) there is a sharp transition from $w_2=-0.95$ to $w_2=-0.8$ (Model II) or to $w_2=-0.6$ (Model III) at $t=0.4 \, t_{cI}$. \label{fig-1}}
\end{figure}

\subsection{Inhomogeneous dark energy \label{inho}}

We shall now consider the possibility that the DE density is inhomogeneous ($\rho_{w1} \neq \rho_{w2}$). For simplicity, we start by assuming that the DE is characterized by a time-independent $w$ (see \cite{Avelino:2009ze,Avelino:2011ey} for a discussion of quintessence and tachyon DE models  with a constant equation of state parameter). In this case
\be
\Delta f=-\frac{4\pi G}{3}  \rho_{w1} (1+3w) \delta_w=-\frac{H_1^2}{2} \Omega_{w1} (1+3w) \delta_w \,,
\label{df}
\ee
with $\delta_w\equiv (\rho_{w2}-\rho_{w1})/\rho_{w1} \neq 0$. If $w=-1/3$ then $\Delta f=0$ and, consequently, the standard LI equation is again recovered. However, in general, Eq. (\ref{df}) leads to a time-dependent correction to the standard LI equation with
\be
\alpha=\frac12 \frac{H_1}{\Delta H} \Omega_{w1} (1+3w) \delta_w\,,
\label{alpharef}
\ee
where
\bq
\frac{\Delta H}{H_1}&=&\frac{H_2}{H_1}-1=\pm {\sqrt{\frac{\Omega_{m1} \rho_{m2}}{\Omega_{m2}\rho_{m1}}}}-1= \nonumber\\
&=&\pm {\sqrt{\frac{\Omega_{m1}}{\Omega_{m2}}\left(1+\delta\right)}}-1\,.
\label{deltaref}
\eq
Note that $\Omega_{m2} \to 1$ in the $\delta \to 0$ and $\delta \to \infty$ limits.

In this paper we shall consider models with a negligible sound speed ($c_s=0$), for which the impact of DE perturbations is expected to be maximum (excluding models with imaginary sound speeds). In these models the DE component remains comoving with the CDM \cite{Creminelli:2009mu}, thus collapsing along with it so that $\rho_{w1} \propto a_1^{-3(w+1)}$ and $\rho_{w2} \propto a_2^{-3(w+1)}$. Using Eq. (\ref{delta}) one finds
\be
\delta_w = \left(\frac{a_1}{a_2}\right)^{3(w+1)}-1 = \left(\delta+1\right)^{w+1}-1\,.
\ee
Expanding around $w\sim -1$ one obtains
\be
\delta_w \sim (1+w) \ln \left(\delta+1\right)\,.
\label{deltaw}
\ee
As expected, using Eqs. (\ref{alpharef}), (\ref{deltaref}) and (\ref{deltaw}), we find that $\alpha \to 0$ in the $w \to -1$, $\delta \to 0$ and $\delta \to \infty$ limits (in the later case, assuming that $w < -0.5$). The fact that the corrections to the standard LI equation vanish in the low and high density perturbation limits implies that the largest corrections to the standard LI equation are expected to occur for objects which are only mildly non-linear, such as clusters of galaxies.

In order to better quantify the modifications to the standard LI equation which arise in the presence of DE perturbations, we show in Fig. \ref{fig-1} the evolution of $\alpha$ with $t/t_{cI}$ ($t_{cI}$ is the perturbation collapsing time in Model I) for three different models. For the background evolution of the various models we consider a fixed value of the DE equation of state parameter ($w=-0.95$) compatible with the latest observational data \cite{Ade:2013zuv} and assume that $t_c$ coincides with the present age of the Universe $t_0$, with $\Omega_{w10}=0.7$ (also in agreement with \cite{Ade:2013zuv}).  Model I (solid line) has fixed value of $w=-0.95$ in the background and perturbed regions and, in this case, the value of $\alpha$ is never very large ($\alpha$ is always smaller than 0.04). For other choices of $w$ close to $-1$ the results would scale roughly with $w+1$.

In models II and III we consider the possibility that the value of $w$ inside the collapsing region becomes different from the background value. In Models II (dashed line) and III (dot-dashed line) we include a sharp transition (at $t=0.4 \, t_{cI}$) of the value of the dark equation of state parameter in perturbed region from $w_2=-0.95$ to $w_2=-0.8$ (Model II) or to $w_2=-0.6$ (model III). As expected, the perturbation collapsing time is not the same for all the models, being a decreasing function of $w_2$. In Models II and III the corrections to the standard LI equation can be much larger than in Model I, thus reflecting a significant impact of the DE perturbations on the dynamics of large cosmological structures. Although the modeling of the role of DE perturbations in the formation and evolution of realistic cosmological structures is outside the scope of the present paper, the maximum variation of $\alpha$ obtained for each model, using the spherical collapse model, is expected to constitute a conservative upper limit to the magnitude of the effect of DE perturbations on the dynamics of collapsed objects such as clusters of galaxies. 

\section{Conclusions \label{conc}}

In this paper we generalized (in the framework of the spherical collapse model) the standard Layzer-Irvine equation to account for DE perturbations. We have quantified the corrections with respect to the standard case, showing that these are expected to be small for models with a constant DE equation of state parameter consistent with the latest observational data, even if the DE has a negligible sound speed. Still, we have shown that much larger corrections may be expected in models with a substantial variation of the DE equation of state parameter between the perturbed and background regions. Although our results were obtained in the context of the spherical collapse model, they allow us to estimate the maximum impact that DE  perturbations can have on the dynamics of clusters of galaxies, which will be probed by ESA's Euclid mission \cite{Amendola:2012ys} with unprecedented precision. This work also provides an important tool which may be used to test the accuracy of a new generation of N-body and Hydrodynamical codes incorporating DE perturbations.

\begin{acknowledgments}
P. A. is partially supported through the project PTDC/FIS/111725/2009 (FCT-Portugal). C. G. is supported through the Incentive 
to Research Program of the Gulbenkian Foundation (Portugal).
\end{acknowledgments}


\bibliography{LI}

\begin{thebibliography}{30}
\expandafter\ifx\csname natexlab\endcsname\relax\def\natexlab#1{#1}\fi
\expandafter\ifx\csname bibnamefont\endcsname\relax
  \def\bibnamefont#1{#1}\fi
\expandafter\ifx\csname bibfnamefont\endcsname\relax
  \def\bibfnamefont#1{#1}\fi
\expandafter\ifx\csname citenamefont\endcsname\relax
  \def\citenamefont#1{#1}\fi
\expandafter\ifx\csname url\endcsname\relax
  \def\url#1{\texttt{#1}}\fi
\expandafter\ifx\csname urlprefix\endcsname\relax\def\urlprefix{URL }\fi
\providecommand{\bibinfo}[2]{#2}
\providecommand{\eprint}[2][]{\url{#2}}

\bibitem[{\citenamefont{{Irvine}}(1961)}]{Irvine:1961}
\bibinfo{author}{\bibfnamefont{W.~M.} \bibnamefont{{Irvine}}}, Ph.D. thesis,
  \bibinfo{school}{HARVARD UNIVERSITY.} (\bibinfo{year}{1961}).

\bibitem[{\citenamefont{{Layzer}}(1963)}]{Layzer:1963}
\bibinfo{author}{\bibfnamefont{D.}~\bibnamefont{{Layzer}}},
  \bibinfo{journal}{\apj} \textbf{\bibinfo{volume}{138}}, \bibinfo{pages}{174}
  (\bibinfo{year}{1963}).

\bibitem[{\citenamefont{Davis et~al.}(1997)\citenamefont{Davis, Miller, and
  White}}]{Davis:1997vg}
\bibinfo{author}{\bibfnamefont{M.}~\bibnamefont{Davis}},
  \bibinfo{author}{\bibfnamefont{A.}~\bibnamefont{Miller}}, \bibnamefont{and}
  \bibinfo{author}{\bibfnamefont{S.~D.~M.} \bibnamefont{White}},
  \bibinfo{journal}{Astrophys. J.} \textbf{\bibinfo{volume}{490}},
  \bibinfo{pages}{63} (\bibinfo{year}{1997}).

\bibitem[{\citenamefont{Liddle and Lyth}(2000)}]{Liddle}
\bibinfo{author}{\bibfnamefont{A.~R.} \bibnamefont{Liddle}} \bibnamefont{and}
  \bibinfo{author}{\bibfnamefont{D.~H.} \bibnamefont{Lyth}},
  \emph{\bibinfo{title}{{Cosmological Inflation and Large-Scale Structure}}}
  (\bibinfo{publisher}{{Cambridge University Press}}, \bibinfo{year}{2000}).

\bibitem[{\citenamefont{Fukugita and Peebles}(2004)}]{Fukugita:2004ee}
\bibinfo{author}{\bibfnamefont{M.}~\bibnamefont{Fukugita}} \bibnamefont{and}
  \bibinfo{author}{\bibfnamefont{P.~E.} \bibnamefont{Peebles}},
  \bibinfo{journal}{Astrophys.J.} \textbf{\bibinfo{volume}{616}},
  \bibinfo{pages}{643} (\bibinfo{year}{2004}).

\bibitem[{\citenamefont{Zaroubi and Branchini}(2005)}]{Zaroubi:2004vu}
\bibinfo{author}{\bibfnamefont{S.}~\bibnamefont{Zaroubi}} \bibnamefont{and}
  \bibinfo{author}{\bibfnamefont{E.}~\bibnamefont{Branchini}},
  \bibinfo{journal}{Mon. Not. Roy. Astron. Soc.}
  \textbf{\bibinfo{volume}{357}}, \bibinfo{pages}{527} (\bibinfo{year}{2005}).

\bibitem[{\citenamefont{{Joyce} and {Labini}}(2013)}]{Joyce:2013}
\bibinfo{author}{\bibfnamefont{M.}~\bibnamefont{{Joyce}}} \bibnamefont{and}
  \bibinfo{author}{\bibfnamefont{F.~S.} \bibnamefont{{Labini}}},
  \bibinfo{journal}{\mnras} \textbf{\bibinfo{volume}{429}},
  \bibinfo{pages}{1088} (\bibinfo{year}{2013}).

\bibitem[{\citenamefont{{Labini}}(2013)}]{Labini:2013}
\bibinfo{author}{\bibfnamefont{F.~S.} \bibnamefont{{Labini}}},
  \bibinfo{journal}{\aap} \textbf{\bibinfo{volume}{552}}, \bibinfo{pages}{A36}
  (\bibinfo{year}{2013}).

\bibitem[{\citenamefont{Bertolami et~al.}(2007)\citenamefont{Bertolami,
  Gil~Pedro, and Le~Delliou}}]{Bertolami:2007zm}
\bibinfo{author}{\bibfnamefont{O.}~\bibnamefont{Bertolami}},
  \bibinfo{author}{\bibfnamefont{F.}~\bibnamefont{Gil~Pedro}},
  \bibnamefont{and}
  \bibinfo{author}{\bibfnamefont{M.}~\bibnamefont{Le~Delliou}},
  \bibinfo{journal}{Phys.Lett.} \textbf{\bibinfo{volume}{B654}},
  \bibinfo{pages}{165} (\bibinfo{year}{2007}).

\bibitem[{\citenamefont{Abdalla et~al.}(2009)\citenamefont{Abdalla, Abramo,
  Sodre, and Wang}}]{Abdalla:2007rd}
\bibinfo{author}{\bibfnamefont{E.}~\bibnamefont{Abdalla}},
  \bibinfo{author}{\bibfnamefont{L.~R.~W.} \bibnamefont{Abramo}},
  \bibinfo{author}{\bibfnamefont{L.}~\bibnamefont{Sodre}}, \bibnamefont{and}
  \bibinfo{author}{\bibfnamefont{B.}~\bibnamefont{Wang}},
  \bibinfo{journal}{Phys.Lett.} \textbf{\bibinfo{volume}{B673}},
  \bibinfo{pages}{107} (\bibinfo{year}{2009}).

\bibitem[{\citenamefont{Abdalla et~al.}(2010)\citenamefont{Abdalla, Abramo, and
  de~Souza}}]{Abdalla:2009mt}
\bibinfo{author}{\bibfnamefont{E.}~\bibnamefont{Abdalla}},
  \bibinfo{author}{\bibfnamefont{L.~R.} \bibnamefont{Abramo}},
  \bibnamefont{and} \bibinfo{author}{\bibfnamefont{J.~C.}
  \bibnamefont{de~Souza}}, \bibinfo{journal}{Phys.Rev.}
  \textbf{\bibinfo{volume}{D82}}, \bibinfo{pages}{023508}
  (\bibinfo{year}{2010}).

\bibitem[{\citenamefont{Bertolami et~al.}(2012)\citenamefont{Bertolami,
  Gil~Pedro, and Le~Delliou}}]{Bertolami:2012yp}
\bibinfo{author}{\bibfnamefont{O.}~\bibnamefont{Bertolami}},
  \bibinfo{author}{\bibfnamefont{F.}~\bibnamefont{Gil~Pedro}},
  \bibnamefont{and}
  \bibinfo{author}{\bibfnamefont{M.}~\bibnamefont{Le~Delliou}},
  \bibinfo{journal}{Gen.Rel.Grav.} \textbf{\bibinfo{volume}{44}},
  \bibinfo{pages}{1073} (\bibinfo{year}{2012}).

\bibitem[{\citenamefont{Avelino and Barreira}(2012)}]{Avelino:2011zx}
\bibinfo{author}{\bibfnamefont{P.~P.} \bibnamefont{Avelino}} \bibnamefont{and}
  \bibinfo{author}{\bibfnamefont{A.}~\bibnamefont{Barreira}},
  \bibinfo{journal}{Phys.Rev.} \textbf{\bibinfo{volume}{D85}},
  \bibinfo{pages}{063504} (\bibinfo{year}{2012}).

\bibitem[{\citenamefont{Shtanov and Sahni}(2010)}]{Shtanov:2010iy}
\bibinfo{author}{\bibfnamefont{Y.}~\bibnamefont{Shtanov}} \bibnamefont{and}
  \bibinfo{author}{\bibfnamefont{V.}~\bibnamefont{Sahni}},
  \bibinfo{journal}{Phys.Rev.} \textbf{\bibinfo{volume}{D82}},
  \bibinfo{pages}{101503} (\bibinfo{year}{2010}).

\bibitem[{\citenamefont{Mota et~al.}(2008)\citenamefont{Mota, Shaw, and
  Silk}}]{Mota:2007zn}
\bibinfo{author}{\bibfnamefont{D.~F.} \bibnamefont{Mota}},
  \bibinfo{author}{\bibfnamefont{D.~J.} \bibnamefont{Shaw}}, \bibnamefont{and}
  \bibinfo{author}{\bibfnamefont{J.}~\bibnamefont{Silk}},
  \bibinfo{journal}{Astrophys.J.} \textbf{\bibinfo{volume}{675}},
  \bibinfo{pages}{29} (\bibinfo{year}{2008}).

\bibitem[{\citenamefont{Avelino et~al.}(2008)\citenamefont{Avelino, Beca, and
  Martins}}]{Avelino:2008cu}
\bibinfo{author}{\bibfnamefont{P.~P.} \bibnamefont{Avelino}},
  \bibinfo{author}{\bibfnamefont{L.~M.~G.} \bibnamefont{Beca}},
  \bibnamefont{and} \bibinfo{author}{\bibfnamefont{C.~J. A.~P.}
  \bibnamefont{Martins}}, \bibinfo{journal}{Phys.Rev.}
  \textbf{\bibinfo{volume}{D77}}, \bibinfo{pages}{101302}
  (\bibinfo{year}{2008}).

\bibitem[{\citenamefont{Bean and Dore}(2004)}]{Bean:2003fb}
\bibinfo{author}{\bibfnamefont{R.}~\bibnamefont{Bean}} \bibnamefont{and}
  \bibinfo{author}{\bibfnamefont{O.}~\bibnamefont{Dore}},
  \bibinfo{journal}{Phys.Rev.} \textbf{\bibinfo{volume}{D69}},
  \bibinfo{pages}{083503} (\bibinfo{year}{2004}).

\bibitem[{\citenamefont{Nunes and Mota}(2006)}]{Nunes:2004wn}
\bibinfo{author}{\bibfnamefont{N.~J.} \bibnamefont{Nunes}} \bibnamefont{and}
  \bibinfo{author}{\bibfnamefont{D.~F.} \bibnamefont{Mota}},
  \bibinfo{journal}{Mon.Not.Roy.Astron.Soc.} \textbf{\bibinfo{volume}{368}},
  \bibinfo{pages}{751} (\bibinfo{year}{2006}).

\bibitem[{\citenamefont{Nunes et~al.}(2006)\citenamefont{Nunes, da~Silva, and
  Aghanim}}]{Nunes:2005fn}
\bibinfo{author}{\bibfnamefont{N.~J.} \bibnamefont{Nunes}},
  \bibinfo{author}{\bibfnamefont{A.~C.} \bibnamefont{da~Silva}},
  \bibnamefont{and} \bibinfo{author}{\bibfnamefont{N.}~\bibnamefont{Aghanim}},
  \bibinfo{journal}{Astron.Astrophys.} \textbf{\bibinfo{volume}{450}},
  \bibinfo{pages}{899} (\bibinfo{year}{2006}).

\bibitem[{\citenamefont{Abramo et~al.}(2007)\citenamefont{Abramo, Batista,
  Liberato, and Rosenfeld}}]{Abramo:2007iu}
\bibinfo{author}{\bibfnamefont{L.}~\bibnamefont{Abramo}},
  \bibinfo{author}{\bibfnamefont{R.}~\bibnamefont{Batista}},
  \bibinfo{author}{\bibfnamefont{L.}~\bibnamefont{Liberato}}, \bibnamefont{and}
  \bibinfo{author}{\bibfnamefont{R.}~\bibnamefont{Rosenfeld}},
  \bibinfo{journal}{JCAP} \textbf{\bibinfo{volume}{0711}}, \bibinfo{pages}{012}
  (\bibinfo{year}{2007}).

\bibitem[{\citenamefont{Ballesteros and Riotto}(2008)}]{Ballesteros:2008qk}
\bibinfo{author}{\bibfnamefont{G.}~\bibnamefont{Ballesteros}} \bibnamefont{and}
  \bibinfo{author}{\bibfnamefont{A.}~\bibnamefont{Riotto}},
  \bibinfo{journal}{Phys.Lett.} \textbf{\bibinfo{volume}{B668}},
  \bibinfo{pages}{171} (\bibinfo{year}{2008}).

\bibitem[{\citenamefont{Abramo et~al.}(2009)\citenamefont{Abramo, Batista, and
  Rosenfeld}}]{Abramo:2009ne}
\bibinfo{author}{\bibfnamefont{L.~R.} \bibnamefont{Abramo}},
  \bibinfo{author}{\bibfnamefont{R.~C.} \bibnamefont{Batista}},
  \bibnamefont{and}
  \bibinfo{author}{\bibfnamefont{R.}~\bibnamefont{Rosenfeld}},
  \bibinfo{journal}{JCAP} \textbf{\bibinfo{volume}{0907}}, \bibinfo{pages}{040}
  (\bibinfo{year}{2009}).

\bibitem[{\citenamefont{Creminelli et~al.}(2010)\citenamefont{Creminelli,
  D'Amico, Norena, Senatore, and Vernizzi}}]{Creminelli:2009mu}
\bibinfo{author}{\bibfnamefont{P.}~\bibnamefont{Creminelli}},
  \bibinfo{author}{\bibfnamefont{G.}~\bibnamefont{D'Amico}},
  \bibinfo{author}{\bibfnamefont{J.}~\bibnamefont{Norena}},
  \bibinfo{author}{\bibfnamefont{L.}~\bibnamefont{Senatore}}, \bibnamefont{and}
  \bibinfo{author}{\bibfnamefont{F.}~\bibnamefont{Vernizzi}},
  \bibinfo{journal}{JCAP} \textbf{\bibinfo{volume}{1003}}, \bibinfo{pages}{027}
  (\bibinfo{year}{2010}).

\bibitem[{\citenamefont{de~Putter et~al.}(2010)\citenamefont{de~Putter,
  Huterer, and Linder}}]{dePutter:2010vy}
\bibinfo{author}{\bibfnamefont{R.}~\bibnamefont{de~Putter}},
  \bibinfo{author}{\bibfnamefont{D.}~\bibnamefont{Huterer}}, \bibnamefont{and}
  \bibinfo{author}{\bibfnamefont{E.~V.} \bibnamefont{Linder}},
  \bibinfo{journal}{Phys.Rev.} \textbf{\bibinfo{volume}{D81}},
  \bibinfo{pages}{103513} (\bibinfo{year}{2010}).

\bibitem[{\citenamefont{Blomqvist et~al.}(2010)\citenamefont{Blomqvist,
  Enander, and Mortsell}}]{Blomqvist:2010ky}
\bibinfo{author}{\bibfnamefont{M.}~\bibnamefont{Blomqvist}},
  \bibinfo{author}{\bibfnamefont{J.}~\bibnamefont{Enander}}, \bibnamefont{and}
  \bibinfo{author}{\bibfnamefont{E.}~\bibnamefont{Mortsell}},
  \bibinfo{journal}{JCAP} \textbf{\bibinfo{volume}{1010}}, \bibinfo{pages}{018}
  (\bibinfo{year}{2010}).

\bibitem[{\citenamefont{Peebles}(1993)}]{Peebles}
\bibinfo{author}{\bibfnamefont{P.~J.} \bibnamefont{Peebles}},
  \emph{\bibinfo{title}{{Principles of Physical Cosmology}}}
  (\bibinfo{publisher}{Princeton University Press}, \bibinfo{year}{1993}).

\bibitem[{\citenamefont{Avelino et~al.}(2009)\citenamefont{Avelino, Trindade,
  and Viana}}]{Avelino:2009ze}
\bibinfo{author}{\bibfnamefont{P.~P.} \bibnamefont{Avelino}},
  \bibinfo{author}{\bibfnamefont{A.~M.~M.} \bibnamefont{Trindade}},
  \bibnamefont{and} \bibinfo{author}{\bibfnamefont{P.~T.~P.}
  \bibnamefont{Viana}}, \bibinfo{journal}{Phys.Rev.}
  \textbf{\bibinfo{volume}{D80}}, \bibinfo{pages}{067302}
  (\bibinfo{year}{2009}).

\bibitem[{\citenamefont{Avelino et~al.}(2011)\citenamefont{Avelino, Losano, and
  Rodrigues}}]{Avelino:2011ey}
\bibinfo{author}{\bibfnamefont{P.~P.} \bibnamefont{Avelino}},
  \bibinfo{author}{\bibfnamefont{L.}~\bibnamefont{Losano}}, \bibnamefont{and}
  \bibinfo{author}{\bibfnamefont{J.~J.} \bibnamefont{Rodrigues}},
  \bibinfo{journal}{Phys.Lett.} \textbf{\bibinfo{volume}{B699}},
  \bibinfo{pages}{10} (\bibinfo{year}{2011}).

\bibitem[{\citenamefont{Ade et~al.}(2013)}]{Ade:2013zuv}
\bibinfo{author}{\bibfnamefont{P.~A.~R.} \bibnamefont{Ade}}
  \bibnamefont{et~al.} (\bibinfo{collaboration}{Planck Collaboration})
  (\bibinfo{year}{2013}), \eprint{1303.5076}.

\bibitem[{\citenamefont{Amendola et~al.}(2012)}]{Amendola:2012ys}
\bibinfo{author}{\bibfnamefont{L.}~\bibnamefont{Amendola}} \bibnamefont{et~al.}
  (\bibinfo{collaboration}{Euclid Theory Working Group})
  (\bibinfo{year}{2012}), \eprint{1206.1225}.

\end{thebibliography}

\end{document}